# CAFE: A Novel Code-switching Dataset for Algerian Dialect, French, and English

LACHEMAT HOUSSAM EDDINE-OTHMAN, LIM Laboratory, University of Bouira, Algeria
ABBAS AKLI, LIM Laboratory, University of Bouira, Algeria
OUKAS NOURREDINE, LIM Laboratory, University of Bouira, Algeria
EL KHEIR YASSINE, German Research Center for Artificial Intelligence (DFKI), Germany
HABOUSSI SAMIA, LIM Laboratory, University of Bouira, Algeria
SHAMMUR ABSAR CHOWDHURY, Qatar Computing and Research Institute (HBKU), Qatar

The paper introduces and publicly releases[1] CAFE – the first Code-switching dataset between Algerian dialect, French, and English languages. The CAFE speech data is unique for *(a)* its spontaneous speaking style in vivo human-human conversation capturing phenomena like code-switching and overlapping speech, *(b)* addresses distinct linguistic challenges in North African Arabic dialect; *(c)* the CAFE captures dialectal variations from various parts of Algeria within different sociolinguistic contexts. CAFE data contains ≈ 37 hours of speech, with a subset, CAFE-small, of 2 hours and 36 minutes released with manual human annotation including speech segmentation, transcription, explicit annotation of code-switching points, overlapping speech, and other events such as noises, and laughter among others. The rest ≈ 34.58 hours contain pseudo label transcriptions. In addition to the data release, the paper also highlighted the challenges of using state-of-the-art Automatic Speech Recognition (ASR) models such as Whisper large-v{2, 3} and PromptingWhisper to handle such content. Following, we benchmark CAFE data with the aforementioned Whisper models and show how well-designed data processing pipelines and advanced decoding techniques can improve the ASR performance in terms of Mixed Error Rate (MER) of 0.310, Character Error Rate (CER) of 0.329 and Word Error Rate (WER) of 0.538.

---

[1]Data download link available after acceptance.

---

Authors' Contact Information: LACHEMAT Houssam Eddine-Othman, h.lachemat@univ-bouira.dz, LIM Laboratory, University of Bouira, Bouira, Bouira, Algeria; Abbas Akli, LIM Laboratory, University of Bouira, Bouira, Bouira, Algeria; OUKAS Nourredine, LIM Laboratory, University of Bouira, Bouira, Bouira, Algeria; EL KHEIR YASSINE, German Research Center for Artificial Intelligence (DFKI), Berlin, Germany; HABOUSSI SAMIA, LIM Laboratory, University of Bouira, Bouira, Bouira, Algeria; SHAMMUR Absar Chowdhury, Qatar Computing and Research Institute (HBKU), Doha, Qatar.

---

## 1 Introduction

In multilingual regions, code-switching—the alternation between two or more languages within a conversation—poses significant challenges for automatic speech recognition (ASR) systems. Algeria, characterized by its linguistic diversity, offers a unique context where Algerian dialect, French, and English frequently intermingle. The complexity of such linguistic environments necessitates the development of specialized datasets and models to capture and process code-switched speech accurately.

The Algerian dialect, known as Darja, presents unique challenges for ASR due to its distinctive phonological and morphological features, influenced by a mix of Berber, French, Spanish, and Turkish [20]. Phonologically, it is marked by the elision of vowels and consonants, creating a rhythm and pace distinct from other Arabic dialects [20, 29]. Additionally, its simplified grammatical structure and significant incorporation of French and English vocabulary further complicate the ASR process.

Despite advancements in ASR technology, existing systems often struggle with the nuanced dynamics of code-switching, leading to high error rates and reduced performance. There are several datasets available that include Algerian speech, such as the ALGSAD Corpus [29], KALAM'DZ Corpus [10], and others (see section 3 for the available datasets for Algerian Arabic). However, these datasets either do not specifically target code-switching phenomena, lack annotations for code-switching, or focus primarily on read rather than spontaneous speech.

This paper introduces the CAFE (Code-switching Algerian, French, and English) dataset, a novel resource designed to address these challenges by providing a comprehensive collection of spontaneous speech that reflects the real-world linguistic landscape of Algeria. The CAFE dataset stands out due to its inclusion of overlapping segments, dialect-level annotation, extensive dataset statistics, and the capture of various sociolinguistic contexts, making it an invaluable resource for ASR and computational linguistics research. The dataset includes around 35 hours of transcriptions, with 2.52 hours manually annotated (CAFE-small) and 34.58 hours pseudo-labeled (CAFE-large), all of which are publicly available for further research and development.

Furthermore, we also benchmark the CAFE data with existing state-of-the-art Whisper ASR models to explore the challenges introduced by such natural spontaneous data. Our experiments with Whisper-based models [28], including PromptingWhisper [26, 30], and the original Whisper implementation [2] referenced as WhisperOriginal, revealed significant challenges in handling code-switched speech. Our initial benchmarking efforts with PromptingWhisper using the large-v2 Whisper model resulted in a high Mixed Error Rate (MER) and Character Error Rate (CER) of 0.73. By upgrading to the Whisper large-v3 model and incorporating bilingual and multilingual prompting techniques, we improved performance, reducing MER to 0.64 and CER to 0.66. Further refinement with advanced decoding techniques from WhisperOriginal [2], a preprocessing pipeline, and temperature adjustments led to an MER of 0.310, CER of 0.329, and WER of 0.538. The sophisticated decoding mechanisms of WhisperOriginal [2], including temperature-controlled greedy search and fallback strategies, proved effective, reducing MER from 0.73 to 0.34. Additionally, following the benchmarking and analysis, we developed an ASR framework with a data preprocessing pipeline to prepare the pseudo-labeled dataset (CAFE-large). CAFE-large includes around 35 hours of transcriptions, which are publicly available and currently undergoing manual review using our dedicated website to enhance their quality (review Section 7 for details).

Therefore, the contributions are:

---

[2]https://github.com/openai/whisper

(1) Design and Develop the CAFE dataset that encompasses spontaneous Algerian dialectal speech code-switched between English and French languages. A subset manually annotated with speech segmentation, transcription, explicit code-switching, and overlapping speech annotation.
(2) Detailed analysis of the CAFE data.
(3) Benchmarking the CAFE data with Whisper models and exploring how data processing and advanced decoding strategy improves the ASR performance when handling such challenging scenarios.

The remainder of this paper is organized as follows. In section 2, we present a comprehensive linguistic analysis of the Algerian Arabic Dialect and its code-switching phenomena. In Section 3, we review the available datasets for Algerian speech and Arabic code-switching and discuss the recent Whisper-based approaches for handling code-switching. Section 4 details the collection and processing of the CAFE dataset. In Section 5, we present our experimentation and benchmarking results, highlighting the challenges and improvements in ASR performance. Section 6 delves into the analysis of our results, including the impact of preprocessing and decoding strategies. Section 7 discusses the creation of a pseudo-labeled corpus for Algerian code-switching speech. Finally, we conclude the paper in Section 8, summarizing our findings and outlining future work.

## 2 Linguistic Background

The study of Automatic Speech Recognition has made significant strides in various languages, yet the Algerian Arabic Dialect remains under-researched. The Algerian dialect, known as Darja, exhibits a unique linguistic identity distinct from Modern Standard Arabic (MSA) and other regional dialects. This distinctiveness is reflected in its grammar, vocabulary, and pronunciation, shaped by a confluence of influences from Berber, French, Spanish, and Turkish due to Algeria's diverse colonial history [20]. Phonologically, the Algerian Arabic Dialect is characterized by the elision of vowels and consonants, resulting in a distinct rhythm and pace that makes it sound faster and more clipped compared to other Arabic dialects [20, 29]. Its grammatical structure is simplified, marked by the omission of case endings and the use of unique verb conjugations, which contribute to the truncated sentence structures that are a hallmark of the dialect [19, 20]. Algerian Arabic's divergence from other dialects is further illustrated in its phonological and morphological features. Unlike the Gulf Arabic dialect, which aligns closely with MSA and exhibits distinctive verb conjugations and noun pluralizations, the Algerian Arabic Dialect has adopted a simpler grammatical structure with significant phonological modifications. For instance, in some regions of Algeria, the phoneme [q] (ق) is replaced by [k] (ك), such as the word "حق" pronounced as "حك" (meaning "right" in English). In other cities, [q] (ق) is produced as a glottal stop [ʔ] (ء), and [θ] (ث) and [t] (ت) are pronounced as [ts] (تس). Additionally, vocabulary in Algiers includes a significant portion of French words and some English, often with extensive phonological modifications, such as "لاطونسيون" ("la tension," meaning "blood pressure"), "فوتبول" ("football"), "شوبينغ" ("shopping"), and "بيزنس" ("business") [1, 20]. Similarly, Egyptian Arabic, or Masry, although widely recognized due to Egypt's cultural influence, retains more classical features and distinct grammatical rules compared to the Algerian dialect. The Maghrebi dialects, spoken across North African countries, share certain similarities with Algerian Arabic, such as the use of specific negation forms. For instance, in Algerian Arabic, "مانروحش" (I won't go) and in Moroccan Arabic, "مانمشيش" (I won't go), both use the prefix "ما" and the suffix "ش" for negation [9]. However, even within the Maghreb region, there are notable variations in daily phrases and pronunciation. For example, "How are you?" is expressed as "واش راك؟" in Algerian Arabic Dialect and "شني حوالك؟" in Tunisian Arabic.

Algerian Arabic Dialect reveals significant regional variations, with distinct pronunciation patterns and a substantial incorporation of French and English loanwords. Morphologically, the dialect simplifies the rules of written Arabic, using a combination of prefixes and suffixes for verb conjugation and incorporating code-switched words with Arabic affixes and clitics. Examples include "وا decide ي" (they decide) and "وا implement ن" (we implement), showcasing the blend of Arabic morphology with French and English vocabulary [8, 9].

## 3 Related Works

The challenges in data modeling and transcription for Algerian Arabic and code-switching contexts are significant, as detailed in Section 2. Accurate annotation and transcription are complicated by the dialect's distinct phonological and grammatical features. These challenges are further increased by the presence of code-switching, which adds more difficulty in understanding and segmenting multilingual speech. A variety of datasets have been created to enhance the field of ASR, particularly for Arabic code-switching contexts. These resources, alongside recent advancements in ASR technology, are vital for addressing the unique difficulties of multilingual and code-switched speech.

This section reviews the available datasets for the Algerian Arabic Dialect and Arabic code-switching, as well as recent Whisper-based approaches for handling code-switching. A summary of the Algerian Arabic Dialect speech datasets is provided in Table 1.

### 3.1 Datasets

In this section we classify the existing datasets that are related to our study into two classes, the first class presents the Algerian speech datasets and the second one presents the available Arabic code-switching datasets.

- Available Algerian Speech Datasets:
  - *ALGSAD Corpus [29]:* The Algerian-spoken Arabic dialect corpus (ALGSAD) is a significant resource for the study of Algerian Arabic. It consists of recordings from 300 speakers across 11 regions of Algeria, focusing on regional accents in MSA spoken in Algeria. However, it does not specifically target code-switching phenomena and includes read speech rather than spontaneous speech.
  - *KALAM'DZ Corpus [10]:* This corpus contains recordings of the eight main Algerian dialects from TV, radio channels, and YouTube. It includes 4,881 speakers and over 104 hours of audio, offering a rich resource for studying Algerian Arabic Dialect speech patterns. However, only a sample is available online, and it lacks code-switching annotations and clear statistics about the amount of annotated data.
  - *Arabic Dataset for Algerian Tamazight Speakers [24]:* This dataset is specifically designed for recognizing Arabic speech when spoken by Tamazight speakers, broadening the pool of linguistic resources available for research. It addresses the variability of accents and dialects in Arabic speech recognition, particularly focusing on the underrepresented Tamazight-speaking community. However, it does not encompass code-switching, and the collected data was read speech instead of spontaneous speech.
- Available Arabic Code-Switching Datasets
  - *ZAEBUC-Spoken Corpus [17]:* The ZAEBUC-Spoken corpus includes code-switching between MSA, Gulf Arabic, Egyptian Arabic, and English. It was collected through Zoom meetings with speakers from six nationalities, making it valuable for studying bilingual and multilingual speech patterns in spontaneous conversational settings.

- *TunSwitch CS Dataset [2]:* This Tunisian code-switched dataset includes spontaneous speech featuring Tunisian Arabic, French, and English. It was collected from radio broadcasts and podcasts, capturing real-world conversations with significant linguistic mixing. The dataset is publicly available and includes both annotated and unannotated audio samples.
- *FACST Corpus [6]:* This FACST Corpus captures spontaneous code-switching between French and Algerian Arabic Dialect. It includes recordings from 20 bilingual speakers, featuring both read and spontaneous speech. However, this dataset is not publicly available and only includes French-Algerian Arabic Dialect code-switching, which limits its applicability to broader linguistic studies. Additionally, the recordings were made using high-quality equipment, which does not reflect real-world scenarios where audio quality can vary significantly.
- *QASR-Dataset [17, 23]:* The QASR-Dataset is a comprehensive Arabic speech corpus with 2,000 hours of data from Aljazeera broadcasts. It includes multiple Arabic dialects and supports tasks like speech recognition and dialect identification. The dataset also contains segments with code-switching between Arabic, English, and French, where 0.4% of it has code-switching between Arabic and English/French.
- *Saudi Arabic-English Code-Switching [21]:* This dataset contains 89 minutes of speech containing Saudi Arabic-English code-switching through informal dinner gatherings.
- *ArzEn corpus [18]:* The ArzEn corpus is a valuable resource for studying Egyptian Arabic-English code-switching, collected through informal interviews with 38 bilingual Egyptians. It comprises 12 hours of transcribed, validated, and segmented spontaneous speech.
- *ESCWA corpus [14]:* The ESCWA corpus comprises 2.8 hours of audio recordings from United Nations meetings held by the Economic and Social Commission for West Asia (ESCWA). This dataset captures code-switching between Arabic (including various dialects) and English/French.
- *MGB-3 ADI-5 subset [15]:* This dataset consists of a 2-hour annotated subset from the ADI-5 development set used in the MGB-3 challenge [5]. It features Egyptian Arabic and MSA code-switching, annotated at the word level to capture linguistic and acoustic cues.

### 3.2 Whisper-based Approaches for Code-Switching ASR

The Whisper model [28], developed by OpenAI, has demonstrated robust performance in various ASR tasks [26, 28, 30]. However, its application to code-switching speech recognition, especially involving multiple languages within a single utterance, presents unique challenges and opportunities [26, 30]. Several studies have explored adapting Whisper to handle code-switching, each with distinct approaches [4, 26, 30].

*3.2.1 Zero-shot Learning with Prompt Engineering.* One significant approach is zero-shot learning with prompt engineering, as investigated in the study [26]. The researchers did not perform additional training but modified the prompt tokens to include multiple language identifiers (e.g., `<|en|><|fr|>`), leveraging Whisper's existing capabilities to handle multilingual input and guiding it to recognize and transcribe speech that switches between languages. This study found that appropriate prompt engineering could significantly improve Whisper's performance on zero-shot tasks, including code-switching ASR.

*3.2.2 Enhanced Decoding Strategies.* The WhisperOriginal [2], [28] employs multiple heuristics to enhance the decoding process. These strategies include beam search, length normalization, language model fusion, and temperature sampling, which collectively improve the accuracy and robustness of the ASR output.

Table 1. Summary of Key Features and Limitations of Speech Datasets Containing Algerian Arabic Dialect

| Dataset Name | Key Features | Limitations |
| --- | --- | --- |
| ALGSAD Corpus [29] | Recordings from 300 speakers, 11 regions, MSA focus | No code-switching, read speech |
| KALAM'DZ Corpus [10] | 4,881 speakers, 104 hours, 8 main dialects | Only a sample is available online, lacks code-switching annotations, lacks annotation statistics |
| Arabic dataset for Algerian Tamazight speakers [24] | Focus on Tamazight speakers, variability of accents | No code-switching, non-spontaneous speech, limited to underrepresented Tamazight-speaking community |
| ArabAlg [25] | Diverse Arabic speech commands, various domains | No spontaneous speech, no code-switching, focused on smart device commands |
| ZAEBUC-Spoken Corpus [17] | Code-switching in MSA, Gulf Arabic, Egyptian Arabic, and English, spontaneous speech | Not specific to Algerian Arabic Dialect |
| TunSwitch CS Dataset [2] | Spontaneous speech, Tunisian Arabic, French, English, publicly available | Not specific to Algerian Arabic Dialect |
| FACST Corpus [6] | French-Algerian code-switching, spontaneous and read speech | Not publicly available, only French-Algerian code-switching |
| QASR-Dataset [23] | 2,000 hours of multi-dialectal Arabic speech, includes code-switching segments, supports various NLP tasks | 0.4% utterances contains code-switching, uncertain quantity of Algerian Arabic Dialect code-switching |
| ESCWA corpus [14] | 2.8 hours of code-switching between Arabic (various dialects) and English/French | Limited size, not publicly available |

For instance, the WhisperOriginal includes a fallback mechanism within the 'decode_with_fallback' function, which handles cases where the initial decoding might fail or produce poor results by trying different temperatures and falling back to a more robust decoding strategy [28]. This ensures better handling of challenging segments and improves overall transcription accuracy. Temperature controls the randomness of the sampling process during decoding. Lower temperatures produce more deterministic outputs, while higher temperatures introduce diversity [28]. Adjusting the temperature in Whisper helps balance generating accurate transcriptions and exploring alternative possibilities, particularly useful for ambiguous or noisy inputs [22].

In contrast, the study [26] also utilizes both beam search and greedy search for decoding but lacks the robust fallback mechanism and comprehensive configuration options found in the WhisperOriginal. This highlights the potential impact of enhanced decoding strategies and additional fallback methods on ASR performance as discussed in Section 6.

This background sets the stage for further exploration and development of robust ASR techniques capable of handling the intricacies of multilingual and code-switching speech.

### 3.3 Importance and Novelty in CAFE Dataset

While some Arabic and Algerian speech datasets are available, there remains a significant gap in resources that capture the unique nuances of Algerian dialects, particularly in spontaneous and naturalistic settings. Existing datasets such as the ZAEBUC-Spoken corpus [17] focus on a mix of dialects and languages, but they lack dedicated resources for Algerian Arabic Dialect in everyday communication contexts. Similarly, the FACST corpus [6] offers valuable insights into French-Algerian code-switching but is not publicly available and does not cover switching between the Algerian dialect, French, and English. It also fails to encompass other forms of Algerian Arabic dialects or the diverse sociolinguistic environments in Algeria. Also for the QASR-Dataset [23], despite its extensive coverage, the large size and complexity make it challenging to extract specific details on the amount of Algerian Arabic Dialect code-switching and to create a subset for the Algerian dialect code-switching. Attempts to run dialect identification models yielded unsatisfactory results, with many utterances in MSA being misclassified as Algerian dialect. In the future, we plan to leverage our CAFE dataset and other resources to develop a language identification model capable of accurately detecting the Algerian dialect.

Our research aims to address this gap by developing the CAFE dataset, which includes spontaneous code-switching speech in the Algerian Arabic Dialect and captures a wider range of dialectal variations and sociolinguistic contexts. The CAFE dataset features code-switching between Algerian dialect, Arabic, French, and English, collected from YouTube to ensure a variety of audio quality and spontaneous speech. It encompasses different accents and includes contributions from around 100 speakers. Notably, 2.52 hours of the CAFE dataset (CAFE-small) is manually annotated by the authors with one version following the same guidelines as the ZAEBUC-Spoken corpus [17], facilitating the merging and expansion of these resources for comprehensive linguistic and ASR research. and 34.58 hours pseudo-labeled dataset (CAFE-large) was prepared based on the best techniques found in section 6. The CAFE dataset is publicly available, providing a substantial resource for further research and development.

By providing a more nuanced and representative dataset, we aim to improve ASR system performance and contribute to the broader field of NLP, particularly for Algerian dialects and multilingual speech.

## 4 CAFE dataset

### 4.1 Dataset collection

The CAFE dataset was meticulously assembled to ensure diverse and comprehensive coverage, vital for robust ASR for the Algerian dialect. Our decision to source data from YouTube channels, rather than recording sessions, was driven by the benefits of real-world scenarios and spontaneous speech. Unlike read speech, which can be scripted and controlled, spontaneous speech in real-world contexts introduces complexities crucial for developing efficient ASR systems.

Additionally, we aimed to capture spontaneous code-switching speech encompassing Algerian dialect, Arabic, French, and English (see statistics in Tables 2, 3, and 5). This inclusion allows for an in-depth analysis of this phenomenon, exploring aspects such as conjugation patterns during code-switching instances. YouTube content, particularly from podcast channels like the Gusra Podcast [3], provides a rich source of spontaneous speech, reflecting natural communication dynamics. This includes instances of overlapping speech, repetitions, variations in pronunciation, and interruptions, all of which are prevalent in real-world conversations but often absent in scripted recordings. By capturing these nuances, our dataset better simulates the challenges faced by ASR systems in practical usage scenarios.

---

[3]https://www.youtube.com/@Gusra

To extract the audio content from YouTube, we employed the powerful YouTube-dl library [4]. We configured the extraction process to optimize for high-quality audio while maintaining efficiency. This included preferences for the best audio format available, conversion to WAV format with a sample rate of 48kHz, a 16-bit depth, and mono channel, ensuring compatibility and quality for subsequent processing stages.

We collected around 37 hours of audio data and created two subsets. The first subset (CAFE-small), consisting of 2 hours and 36 minutes, was used for experimentation and analysis, as well as for applying further advanced annotation techniques. The second subset (CAFE-large), which comprises the remaining data, was annotated and is currently being manually reviewed.

Our dataset also incorporates non-lexical vocalizations such as laughter, crying, coughing, and filler sounds like "mhm" and "umm". These elements were annotated in CAFE-small using the ZAEBUC-Spoken guidelines [17]. Also inherent to spontaneous speech, it adds further complexity to the ASR task, requiring models to accurately distinguish between meaningful words and non-linguistic vocalizations.

Key among these challenges is the need to accurately transcribe speech despite its inherent variability and unpredictability. Spontaneous speech encompasses diverse linguistic patterns, accents, and dialects, along with code-switching between languages, as observed in our dataset (see Table 9). These complexities are essential for training and benchmarking ASR models across diverse linguistic contexts.

### 4.2 Data processing

In the data processing phase (Figure 1 illustrates the stages of the data crawling and processing pipeline), our objective was to segment the audio files into manageable chunks, each encapsulating complete contextual information crucial for effective ASR model training. Particularly in the era of Large Language Models (LLMs), where the context plays a pivotal role in enhancing Artificial Intelligence (AI) understanding and performance [25, 31]. For instance, in the OpenAI's implementation [2], the context of the previous 30 seconds of the chunk is passed to the decoder to improve performance [25]. Therefore, ensuring that each chunk ends at a point of silence, signifying the completion of speech and context, becomes paramount. This meticulous segmentation not only facilitates more accurate transcription but also empowers ASR models to better comprehend and generate human-like responses by leveraging the broader contextual understanding provided by these LLMs.

The duration of each chunk was carefully calibrated to balance context completeness and processing efficiency. For this, we used pydub split on silence method [5] with minimum silence length equals 1000 milliseconds and silence threshold equals -45db as parameters. We targeted durations ranging from 25 to 60 seconds for the initial segmentation, aiming to encapsulate sufficient contextual information while avoiding overly lengthy segments that may hinder processing speed or introduce computational. This range was informed by previous research findings and best practices in ASR dataset preparation [17], providing a solid foundation for our segmentation strategy.

If the desired number of chunks per audio file (up to 10) was not achieved within this range, a secondary segmentation pass was conducted, targeting durations from 15 to 120 seconds for the remaining chunks. This extended range allowed for more flexibility in accommodating audio files with varying lengths and complexities while maintaining the overarching goal of context completeness.

---

[4] https://github.com/ytdl-org/youtube-dl
[5] https://pydub.com/

Manuscript submitted to ACM

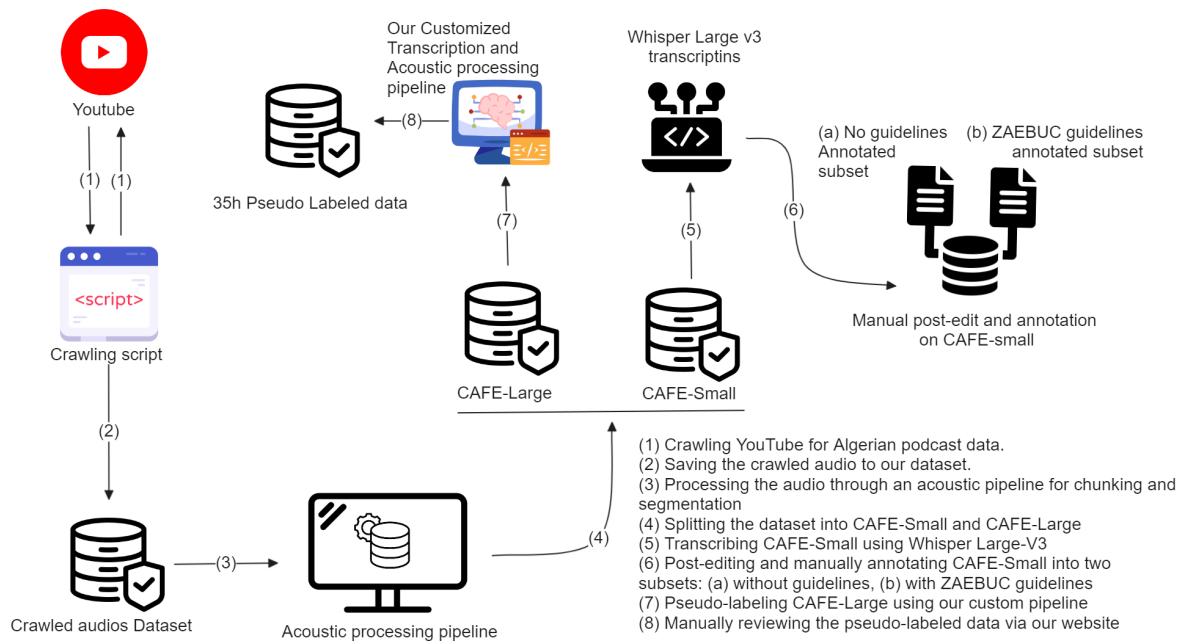

Fig. 1. CAFE Dataset Construction Pipeline.

It's noteworthy that we deliberately retained instances of overlapping speech, background noises, and music in some chunks. This decision was motivated by a desire to create a more realistic and challenging dataset, better suited for evaluating ASR models in real-world scenarios where such complexities are commonplace.

By automating the chunking process and incorporating varied durations and contextual considerations, we ensured that our dataset captures the intricacies of spontaneous speech while providing sufficient context for effective ASR model training in the future. Please refer to Tables 2, 3, and 5 for detailed statistics regarding the processed and annotated corpus.

Table 2. General Details About CAFE-Small Dataset

| Number of voices | 35 (4 females, 31 males) |
|---|---|
| Duration | 2h36mn |
| Size | 857MB |
| Dialectness levels | L0: 10, L1: 11, L2: 14, L3: 79, L4: 57 |

Table 3. Utterance Duration Statistics

| Duration (minutes) | 10 < X < 30 | 30 < X < 60 | 60 < X < 120 |
|---|---|---|---|
| Number of utterances | 57 | 115 | 33 |

### 4.3 Whisper Model for Pseudo-Labeling

Our pseudo-labeling framework leverages the Whisper Large-v3 model [28], which has shown remarkable performance in multilingual ASR.

*4.3.1 CAFE-small.* To prepare the pseudo labels for this subset before manual post-editing, we used Whisper Large-v3 [28] with default parameters and decoding strategies [2]. This subset was utilized for further analysis and experimentation to improve our pseudo-labeling ASR framework (see section 6).

*4.3.2 CAFE-large.* Based on our findings from the analysis and experimentation on CAFE-small, a pseudo labeling framework using Whisper Large-v3 [2] and acoustic processing pipelines based on the "*pyannote speaker-diarization-3.1*" [11, 27] models was created. This framework was used to prepare the pseudo labels for CAFE-large, which are being manually reviewed on our website (see section 7 for details and table 12 for statistics).

### 4.4 Data annotation

The CAFE-small dataset was annotated through a post-editing process, where annotators were instructed to carefully review and correct the automatic transcriptions. If the quality of the automatic transcription was poor, they were required to manually reannotate the segments from scratch to ensure accuracy. This resulted in two datasets for the CAFE-small subset: one with manual transcription without any guidelines and another using the ZAEBUC-Spoken guidelines [17].

*4.4.1 Manual transcriptions of raw text.* In the initial phase of data annotation, we employed a combination of manual post-editing and full manual transcription to ensure the accuracy of the dataset. The primary goal was to produce transcriptions that precisely mirrored the spoken content. For segments where automatic transcriptions were of acceptable quality, annotators conducted meticulous post-editing to refine the output. However, in cases where the automatic transcriptions were of poor quality, annotators re-transcribed the segments from scratch, capturing every word uttered by the speakers without any alterations or corrections. This process also included accurately transcribing instances of code-switching, where multiple languages were seamlessly interwoven within the dialogue.

For code-switching scenarios involving Arabic or Algerian dialects, Arabic script was employed, while Latin script was utilized for French and English. Numbers were transcribed in numerical format to maintain accuracy. No adjustments were made regarding punctuation, background noises, or unclear words. In cases of mispronunciations, the intended word was transcribed regardless of pronunciation. Additionally, repetitions, incomplete words, and morphologically code-switched words were faithfully represented in the transcriptions. For example:

- Repetitions:

  كنا كنا كنا رايحين لل park

- Incomplete words:

  نتلاقاو at the corn

- Morphologically code-switched words:

  واdecide ي, y decide wA, 'Ils décident', 'they decide', in French and English.

*4.4.2 ZAEBUC guidelines.* For the second phase of data annotation, we followed the transcription standards established by the ZAEBUC-spoken corpus creators [17]. These guidelines provided a comprehensive framework for transcription practices across various linguistic and contextual elements.

The ZAEBUC-Spoken guidelines [17] encompassed rules governing punctuation usage, transcription of numbers, handling of background noise and typing sounds, treatment of unclear words and mispronunciations, and annotation of conversational speech elements such as repetitions, interruptions, and non-speech segments. Moreover, specific instructions were provided for code-switching instances, dictating script usage and format for morphological code-switching scenarios.

By adhering to these established standards, we aimed to ensure uniformity and fidelity in transcribing the dataset (see Table 4 for examples). Furthermore, aligning our annotation practices with those of ZAEBUC-Spoken [17] facilitates potential future integration with their dataset, enhancing comparability and analysis across resources in the field of ASR and computational linguistics.

Table 4. Examples of Manual Annotation and ZAEBUC Guideline Application

| Manual annotation | Applying ZAEBUC guideline | Dialectness Level Annotation |
|---|---|---|
| كالرسائل تع بكري أك شايف الرسائل تع بكري كيفاش تخرجهم كل هكذا يتحلوا هكذا هكذاك الـ فاهمني انت كـ developer تخدم الـ user interface تاعك وكذا ... | [noise] كالرسائل تع بكري أك شايف الرسائل تع بكري كيفاش تخرجهم كل هكذا يتحلوا هكذا هكذاك الـ فاهمني؟ انت كـ + developer تخدم ال + user interface تاعك وكذا ... | L3 |
| السلام عليكم ورحمة الله وبركاته مستمعينا الكرام ومستمعاتنا الكريمات مرحبا بكم في حلقة جديدة مع podcast gusra واليوم معانا ظيف مميز وهو ... | [noise] السلام عليكم ورحمة الله وبركاته. مستمعينا الكرام ومستمعاتنا الكريمات مرحبا بكم في حلقة جديدة مع **gusra** podcast [noise]، واليوم معانا ظيف مميز وهو ... | L1 |
| كانوا يشروهم كي تشوف واحد يهدر بـ Téléphone portable بكري اوه اوه راو هايل كان القمقوم الشريحة هذيك تع جيزي كانت مليون وميتين ... | بكري Téléphone portable كانوا يشروهم كي تشوف واحد يهدر بـ %أوه% ** أوه راو هايل كان ** القمقوم الشريحة هذيك تع جيزي كانت مليون وميتين ... | L3 |
| يعاود يقص مباعد بالوقت ولا ذرك ولات بحيث دارولهم millennium ودارولهم الباطل واشتراك وشوف ذرك الانترنت ذرك ولات | يعاود يقص {laugh} مباعد بالوقت ولا ذرك ولات بحيث دارولهم millennium ودارولهم الباطل واشتراك وشوف ذرك الأنترنت ذرك ولات— | L3 |

### 4.5 Dataset Statistics

The CAFE dataset's diversity not only facilitates advancements in ASR research but also opens avenues for cross-linguistic studies, supports low-resource language recognition, and enhances ASR performance in multilingual environments. Moreover, to our knowledge, it stands as the first publicly available dataset capturing the nuances of Algerian dialect, Arabic, French, and English code-switching. This distinction underscores its unique contribution to the field, offering researchers a valuable resource for exploring the intricacies of multilingual communication within the Algerian context.

During the course of our work, we were required to calculate the CESAR metric [30] to determine the code-switching rate of our dataset (refer to Table 5). This calculation was crucial for our analysis. For the comprehensive discussion, please refer to Section 6: Results and Discussion. To provide additional insight into the CAFE-small dataset, we calculated the Code Mixing Index (CMI) for all utterances, resulting in an average CMI of 0.254 (see Figure 2).

Table 5. Distribution of Sentences by calculated Code Switching Rate using CESAR metric in CAFE-small dataset

| CESAR range | Number of utterances |
|---|---|
| under 0.2 | 62 |
| 0.2 to 0.4 | 98 |
| Above 0.4 | 10 |

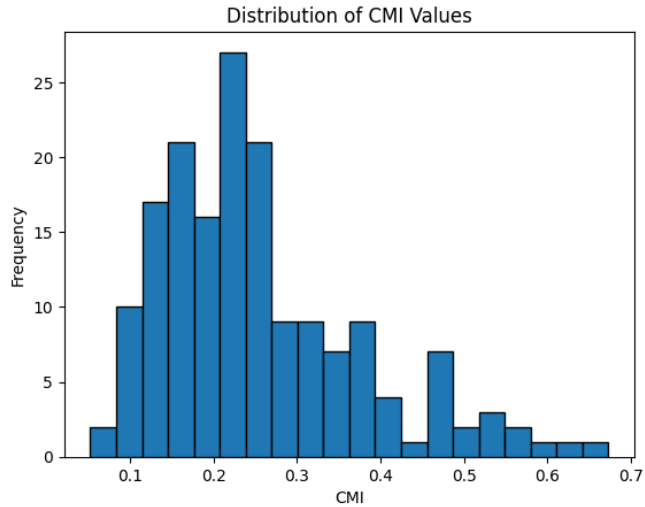

Fig. 2. Distribution of CMI Values of CAFE-small Dataset

We also manually annotated the CAFE-small dataset for dialect levels, using four labels ranging from L0 to L4, as specified in the ZAEBUC-Spoken guidelines [17]. Our dataset is rich in dialect variations, with 136 out of 170 chunks rated above the L3 level .See Table 2.

The CESAR measure is designed to estimate the extent of code-switching in a given corpus C, particularly concerning a reference language r. The CESAR metric is a weighted combination of two parameters as defined in Eq.1. $P_r(C)$ parameter measures the proportion of documents in the corpus that contain code-switching. It assigns a score of 0 if the entire corpus is written in the reference language and 1 if no text is code-switched. The value in between is based on the number of languages present in each document. this is given by the equations ( Eq.2, Eq.3, Eq.4). $B_r(C)$ parameter calculates the rate of "noise" introduced by words that differ from the reference language within each document. It assesses how much of the documents deviate from the reference language, it is given by Eq.5.

$$\text{CESAR}(C) = \alpha P_r(C) + \beta B_r(C) \tag{1}$$

Where:

$$P_r(C) = \frac{1}{n} \sum_{i=1}^{n} \delta(d_i) \text{LF}(d_i) \tag{2}$$

$$\delta(d_i) = \begin{cases} 1 & \text{if } \exists w \in d_i \text{ such that } w \notin \text{reference language} \\ 0 & \text{otherwise} \end{cases} \tag{3}$$

$$\text{LF}(d_i) = \frac{\text{number of distinct languages} - 1}{\text{number of distinct languages}} \tag{4}$$

$$B_r(C) = \frac{1}{n} \sum_{i=1}^{n} \left(\frac{N(w_{d_i})}{N_{d_i}}\right) \text{LF}(d_i) \tag{5}$$

$N(w_{d_i})$ = number of words in $d_i$ not in the ref language

$N_{d_i}$ = total number of words in $d_i$

The weights $\alpha$ and $\beta$ are determined empirically and satisfy $\alpha + \beta = 1$.

### 4.6 Overlapping Speech subset

In our study and analysis 6, we prepared a subset from CAFE-small with a duration of 28 minutes containing 37 out of 170 audio chunks with overlapping speech only, to further evaluate the ASR models. We applied both acoustic and linguistic data processing to this subset, as described below and illustrated in Table 6.

Table 6. Overlapping speech segment examples: before and after removing overlap and linguistic processing.

| Overlapping segments: manual transcription | After Removing Overlapped Segments and Applying Linguistic Processing |
|---|---|
| هي هذه باينة في reddit شغل خارجة خارج صريح أنو جماعة هذه الفكرة تنتمي المجموعة هذي راك حتى راك أثرت نقطة كبيرة ... | هي هذه باينة في reddit شغل خارجة خارج صريح أنو جماعة هذه الفكرة تنتمي حتى راك أثرت نقطة كبيرة ... |
| يعني كان صح صدمة شوية اليامات اللولين. خاصة. هي علي كل حال داخلة فالتكوين اه؟ اه سبحان الله. بصح ثاني خاصة كانت مع رمضان ... | يعني كان صح صدمة شوية اليامات اللولين. خاصة. داخلة فالتكوين اه؟ اه سبحان الله. بصح ثاني خاصة كانت مع رمضان ... |
| par ce que الخيط تاعو تقص انا عندي مليح الخيط تاعو تقص pour une raison donc j'ai préparé quelques diapo ... | par ce que انا عندي مليح الخيط تاعو تقص pour une raison donc j'ai préparé quelques diapo ... |

*4.6.1 Acoustic Processing.* We employed automatic processing to remove the overlapped segments from the audio files. Using the pyannote audio library [11, 27] and the pre-trained model "*pyannote/overlapped-speech-detection*", we detected overlapping speech segments. These segments were then excluded from the audio files to ensure a clean dataset for further analysis. Parameters such as sample rate (48,000 Hz) and data type (PCM 16-bit) were utilized to maintain high audio quality.

*4.6.2 Linguistic Processing.* In the linguistic processing stage, we aligned the transcriptions to fit the new audio files without overlapping segments. This involved removing tokens corresponding to the overlapping speech, as well as any incomplete or unintelligible words. The resulting transcriptions were edited to reflect the cleaned audio, ensuring accurate alignment and readability for further analysis.

## 5 Experimentation and benchmarking

In our research, we explored the capabilities of Whisper-based models for processing our code-switching dataset for the Algerian dialect, French, and English (CAFE-small). Our initial experiments utilized the PromptingWhisper repository, an implementation grounded in the study [26]. This initial phase employed the large-v2 configuration of the Whisper model.

### 5.1 Evaluation metrics

The evaluation metrics used to assess the performance of the ASR systems on the CAFE-small dataset are (1) Word Error Rate (WER), (2) Character Error Rate (CER), and (3) Mixed Error Rate (MER). Each of these metrics provides a different perspective on the accuracy of the transcription, capturing errors at different levels of granularity.

*(1) Word Error Rate* WER is a common metric used to evaluate the performance of ASR systems. It is calculated as the number of errors (substitutions, deletions, and insertions) divided by the total number of words in the reference.

*(2) Character Error Rate* CER is similar to WER but operates at the character level instead of the word level. It is useful for languages and dialects with complex morphology or when word boundaries are not clear.

*(3) Mixed Error Rate* MER is a more refined metric that combines the insights of WER and CER by tokenizing the text in a manner that captures both character-level and word-level errors. This is particularly useful for evaluating ASR performance on dialects with significant phonological and morphological variations. Especially for the text in the Algerian dialect rather than MSA, MER can be more effective in certain situations. For instance, "نقدرو" in Algerian Arabic means "we can," but Whisper might predict "نقدر" which may sound like "نقدرو" but without the letter "و", and the absence of diacritic marks makes it hard to evaluate. The letter "و" signifies the plural "us," so the omission of this single letter can significantly impact the WER, even though the word's meaning is contextually correct. Given the morphological complexities of the dialect, MER provides a more accurate evaluation. For this reason, throughout our research, we will prioritize MER as the primary metric for our analysis.

### 5.2 Whisper-based approaches benchmarking

The initial benchmarking utilized the large-v2 model which is present in the PromptingWhisper original implementation, applying the same prompting technique with bilingual AR-FR and AR-EN prompt. The results from this setup were unsatisfactory and switching between AR-FR and AR-EN didn't impact the results as mentioned in the study [30] as well. Both the Mixed Error Rate (MER) and Character Error Rate (CER), were recorded at 0.73 for both prompts. These results suggested significant limitations in the model's ability to accurately transcribe the multilingual content of the CAFE-small dataset.

Given the suboptimal performance with the large-v2 model, we upgraded to the Whisper large-v3 model while retaining the original decoding and prompting strategies. This adjustment resulted in improved transcription accuracy, with MER and CER dropping to 0.64 and 0.66, respectively, for various bilingual and multilingual prompts (AR-FR, AR-EN, AR-EN-FR, AR-FR-EN). Despite these improvements, the transcription accuracy was still below expectations, and prompting whisper is not improving the task of code-switching and provides similar results to those without prompting which led us to further investigation into the decoding process.

In addition to evaluating MER and CER for all approaches. (see Table 7), we also calculated WER for "WhisperOriginal" and "Preprocessing-V2 with WhisperOriginal" to provide further insights into the results. These calculations were performed for temperature values of 0 (default) and 0.2 (the optimal temperature for CAFE-small based on the MER metric). The results are presented in Table 8. The MER values across various temperatures are visually depicted in Figure 3, highlighting the performance trends of each pipeline.

Table 7. Reported MER and CER for CAFE-small Dataset using different Whisper-based pipelines.

| Pipeline | default (0,0) | | 0,2 | | 0,4 | | 0,6 | | 0,8 | | 1 | |
|---|---|---|---|---|---|---|---|---|---|---|---|---|
| | MER | CER | MER | CER | MER | CER | MER | CER | MER | CER | MER | CER |
| PromptingWhisper - Large v2 | 0.735 | 0.735 | / | / | / | / | / | / | / | / | / | / |
| PromptingWhisper - Large v3 with bilingual and multilingual prompting | 0.643 | 0.665 | / | / | / | / | / | / | / | / | / | / |
| Whisper-Original | 0.333 | 0.345 | 0.335 | 0.352 | 0.35 | 0.367 | 0.366 | 0.381 | 0.449 | 0.465 | 0.624 | 0.652 |
| Prompted-WhisperOriginal | 0.333 | 0.346 | 0.335 | 0.352 | 0.35 | 0.367 | 0.366 | 0.381 | 0.449 | 0.465 | 0.624 | 0.652 |
| Preprocessing-V1 with WhisperOriginal | 0.355 | 0.365 | 0.342 | 0.358 | 0.378 | 0.391 | 0.388 | 0.393 | 0.448 | 0.462 | 0.637 | 0.652 |
| Preprocessing-V2 with WhisperOriginal | 0.316 | 0.339 | 0.31 | 0.329 | 0.343 | 0.357 | 0.349 | 0.363 | 0.432 | 0.444 | 0.632 | 0.649 |

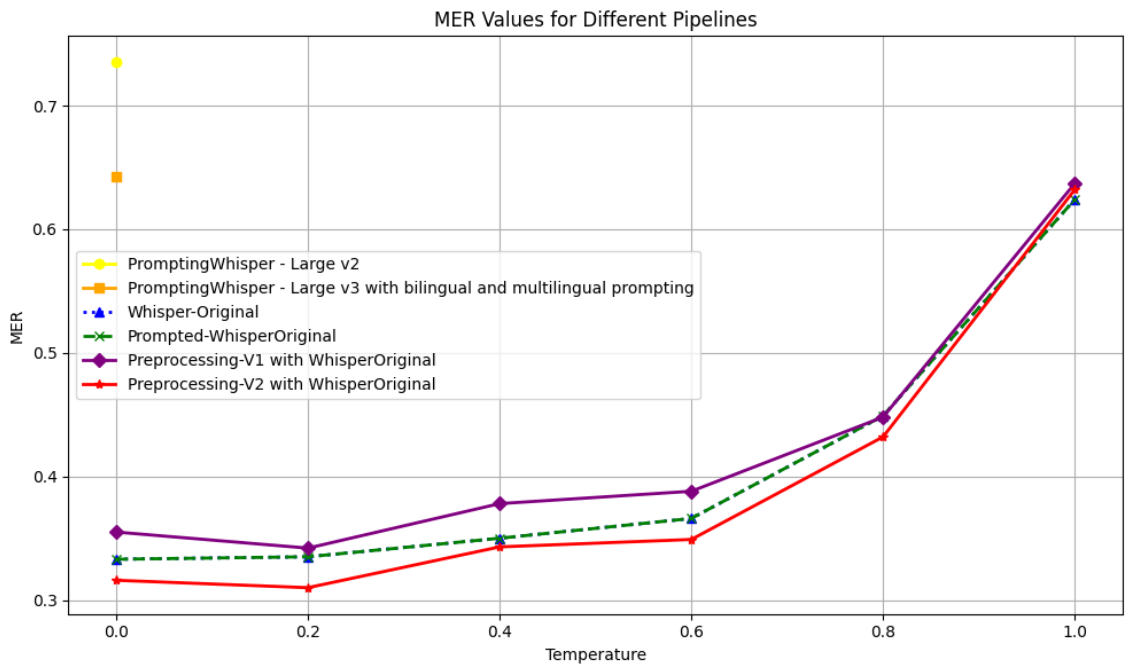

Fig. 3. MER values across different temperature settings for various pipelines.

Table 8. Reported WER, MER, and CER for WhisperOriginal and Preprocessing-V2

| Pipeline | Default (0.0) | | | Temperature = 0.2 | | |
| --- | --- | --- | --- | --- | --- | --- |
| | WER | MER | CER | WER | MER | CER |
| WhisperOriginal | 0.526 | 0.333 | 0.345 | 0.529 | 0.335 | 0.352 |
| Preprocessing-V2 with WhisperOriginal | 0.531 | 0.316 | 0.339 | 0.538 | 0.31 | 0.329 |

## 5.3 Decoding Strategy Analysis

Our investigation into decoding strategies highlights a significant divergence between the WhisperOriginal [2] and the PromptingWhisper repository [6]. TheWhisperOriginal employs a sophisticated approach to decoding that includes mechanisms such as temperature adjustments to optimize transcription accuracy, which was not considered in the PromptingWhisper implementation.

*5.3.1 OpenAI's Advanced Decoding Techniques.* Whisper's decoding process leverages a temperature-controlled greedy search, where the temperature parameter controls the randomness of prediction by adjusting the likelihood of less probable candidates. This technique allows for a more flexible adjustment to transcription challenges, dynamically balancing between accuracy and diversity in the generated output [28].

Additionally, OpenAI's implementation [2] utilizes a fallback strategy when initial predictions do not meet certain quality thresholds, such as when the output is too repetitive or lacks sufficient confidence. This is managed by adjusting several decoding parameters dynamically based on the results of each attempt. Specifically, the fallback mechanism in Whisper adjusts the temperature, beam size, and patience based on real-time output characteristics, allowing for multiple attempts at decoding with varied settings to optimize the result. For example, when the temperature is above zero, beam size and patience are disabled to prioritize quicker, less precise decoding paths. Conversely, at zero temperature, it opts for a strategy focusing on accuracy by disabling the 'best of' setting [2], [28].

*5.3.2 Limitations in PromptingWhisper's Implementation.* In contrast, the PromptingWhisper repository lacks these adaptive decoding enhancements. It does not implement the temperature-controlled greedy search or the dynamic fallback mechanisms found in the WhisperOriginal implementation. This absence is particularly impactful when dealing with the CAFE dataset, where linguistic diversity and complexity can significantly benefit from more nuanced decoding strategies. The lack of these features in PromptingWhisper likely contributes to the lower transcription accuracy observed in our experiments, Because when we benchmarked with WhisperOriginal [2] with the default parameters (temperature set to 0) we got better results, MER and CER were 0. 333, 0. 346 respectively (See Table 7 for benchmarking results).

## 6 Results and Discussion

### 6.1 ASR Performance on CAFE-small

The fallback decoding function plays a crucial role in setting the right parameters, but sometimes it fails and some audio chunk results still face repetitive tokens, and since Whisper is trained on translation task as well, it does the transcription from Algerian dialect, French, or English to the most frequent language in the audio chunk. And on some audio chunks where the reference answer has more than 20 tokens, it predicts few words (see examples in Table 9).

---

[6]https://github.com/jasonppy/PromptingWhisper

This analysis prompted further investigation into the parameters and data, particularly to understand why certain chunks exhibited MER values exceeding 0.60 and did not meet our expectations. the reason behind WhisperOriginal predicting few words in audio chunks that have a good amount of tokens in the reference answer (see Table 9), is because they constrained the initial timestamp token to be between 0.0 and 1.0 second,s and the audio chunks may have non-speech segments such as corrupted audio, laughter, stuttering or overlapping speech segments at the beginning of the audio chunk, and Whisper couldn't capture the first token in that timestamp which leads to hallucination. For instance, when we removed the first 3 seconds in the "*chunk_22.wav*" which contained music, the results improved significantly, reducing the MER from 0.97 to 0.35.

We also sought to determine the most suitable temperature for code-switching ASR for the Algerian dialect, French, and English. To achieve this, we implemented an algorithm that manually tested various temperature values (0.0, 0.2, 0.4, 0.6, 0.8, and 1.0), selecting the optimal result for each segment and recording the corresponding temperature. This approach was motivated by our observation that setting the temperature to 0.2 for the segment "*chunk16v2_6*" significantly improved the result, reducing the MER from 0.571 to 0.249.

Table 9. Analysis of Whisper Transcription Failures for some CAFE-small Audio Chunks

| Reference transcription | Prediction | Observation | Reason |
| --- | --- | --- | --- |
| شكراً جزيلاً سيدعلي يعطيك الصحة بارك الله فيك لاجابة الدعوة وصراحة انا كيما نقولو مزالو ... | شكرا جزيلا | Predicting few words | Chunk started with overlapped speech |
| السلام عليكم ورحمة الله وبركاته مستمعينا الكرام مستمعاتنا الكريمات مرحبا بكم مع حلقة جديدة في podcast gusra واليوم معنا الدكتورة ... | السلام عليكم ورحمة الله وبركاته نبداء مع المواضيع الشيقة التي تتعلق بطب الأسنان مع ضيفتان | Predicting one sentence | Chunk started with introduction music, from 0s to 3s |
| elle était rapide و la réponse elle était claire قالهم bien pensé bien foncé en ce sous و حق أودي مالا كي كنتو نتوما في les entre dames certainement كان الشعب فقير علاه مخممتوش ف la suite كاين ... | la réponse a été claire rapide bien pensée dans un sens où j'ai dit que l'église d'outre-dame à un certain moment était une ville faquée ... | Translation from Arabic to French | CESAR value for this chunk was 0.42 with the Arabic language as reference. The chunk started with French words. |

### 6.2 Code-switching rate of CAFE-small

For deeper insights, we calculated the code-switching rate of the CAFE-small dataset using the CESAR metric [3], with the statistics in Table 5, Then we started the evaluation of WhisperOriginal with different temperature values for each chunk and selected the best value based on the lowest MER value. For the Distribution of Temperature Values for CESAR 0.2 to 0.3 and 0 to 0.6, please refer to Figure 4.

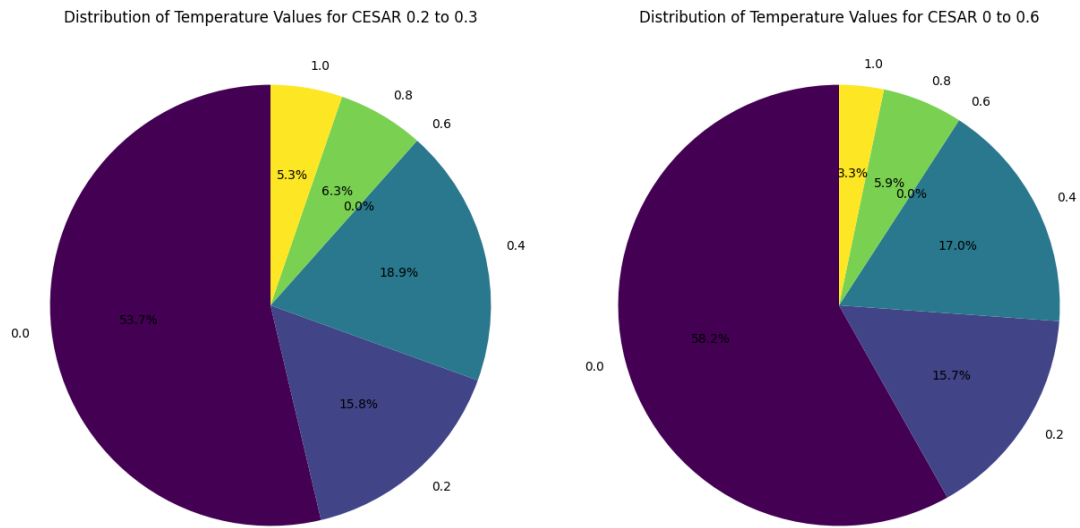

Fig. 4. The Distribution of Temperature Values for CESAR 0.2 to 0.3 and 0.0 to 0.6.

For better insights, we plotted the Correlation between Temperature adjustments and MER in CAFE-small, see Figure 5.

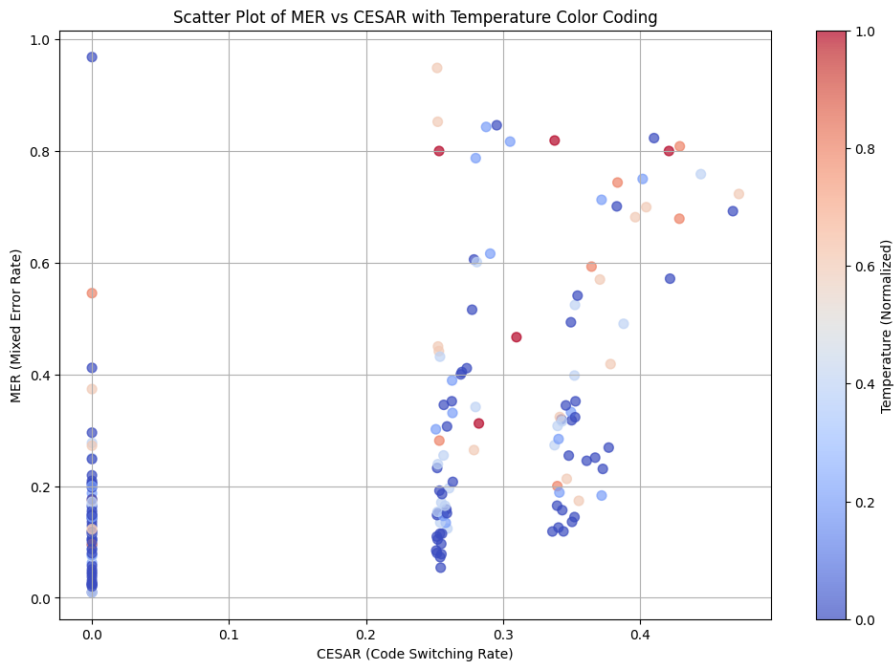

Fig. 5. Correlation between Temperature Adjustments, CESAR Values, and Mixed Error Rates in CAFE-small Dataset.

### 6.3 Data preprocessing experimentation

The preceding analysis and findings prompted us to develop a preprocessing layer to be implemented before Whisper. Additionally, we established a baseline model with fine-tuned parameters for data pseudo-labeling. The results obtained after applying our preprocessing layers are presented in Table 7.

During the development phase of our layer, we tested several data-cleaning strategies. The initial version, referred to as Preprocessing-V1, employed the "*pyannote speaker-diarization-3.1*" [11, 27] and "pyannote overlapped-speech-detection" [12, 13] models to eliminate overlapping speech and non-speech segments such as corrupted audio, laughter, or stuttering lasting longer than 1.5 seconds. The presence of these segments at the beginning of the audio significantly increased the Mixed Error Rate (MER), since the WhisperOriginal was designed to detect the first spoken token within the initial 0 to 1-second interval. The subsequent iteration, Preprocessing-V2, streamlined this process by using only the "pyannote speaker-diarization-3.1" model to remove non-speech segments that exceeded 0.4 seconds

In our evaluation, Preprocessing-V2 demonstrated improved performance by retaining overlapped speech segments, as opposed to Preprocessing-V1, which removed these parts and yielded inferior results. Notably, the outcomes from Preprocessing-V1 were even less favorable than the WhisperOriginal transcription without any preprocessing. During the removal of overlapped speech in Preprocessing-V1, we also excluded the tokens from the utterances within the overlapped period. Since the CAFE-small dataset contains spontaneous speech, removing overlapped speech may corrupt segments by cutting off spoken words, resulting in incomplete tokens and negatively impacting transcription accuracy. The comparative results of these approaches are detailed in Table 7.

To further validate this hypothesis, we created two subsets of the CAFE-small, both containing only the audio chunks with overlapping speech (22% of the CAFE-small dataset). We prepared the two subsets as follows: (a) segments without any overlapping speech (SET1) and (b) segments with overlapping speech (SET2). We then evaluated the performance of the WhisperOriginal on these manually processed chunks using a temperature setting of 0.2. The results showed that the performance on SET2, which contained overlapping speech, was better than on SET1, as shown in Table 10.

Table 10. Performance of Whisper on CAFE-small subsets with and without overlapping speech

|  | CER | MER |
| --- | --- | --- |
| SET1 (without overlapping speech) | 0.436 | 0.410 |
| SET2 (with overlapping speech) | 0.409 | 0.387 |

To validate our hypothesis that the presence of non-speech segments at the beginning of the audio significantly increased the MER, we identified audio chunks with an MER value exceeding 0.80 and manually refined those using Audacity [7]. The refinement process involved eliminating non-speech segments, such as stuttering and silent intervals, especially those present at the beginning of the audio chunks. Subsequently, we evaluated the performance of the WhisperOriginal on these manually processed chunks using two temperature settings, 0.0 and 0.2. The benchmark results for the manually processed chunks were then compared with the unprocessed chunks with the same temperature settings, as detailed in Table 11.

Table 11. Comparison of MER for Processed and Unprocessed Chunks at Different Temperature Settings

|  | Temperature = 0.0 | Temperature = 0.2 |
| --- | --- | --- |
| MER for unprocessed chunks | 0.986 | 0.963 |
| MER for processed chunks | 0.853 | 0.747 |

Although the results improved after applying the preprocessing pipelines to the chunks, we still observe high MER values for certain chunks. This is due to the quality of the recording setup and the presence of corrupted segments within the speech, which disrupt and truncate some words. We conclude that incorporating an advanced preprocessing layer and setting the temperature value to 0.2 can effectively yield acceptable predictions for the CAFE-small dataset. Given our plan to prepare a pseudo-labeled dataset using the Gusra Podcast channel, we will apply the same approach.

## 7 Pseudo labeled corpus for Algerian dialect, French and English code switching

To address the scarcity of data specific to the Algerian code-switched dialect, we have curated a pseudo-labeled dataset from the YouTube channel "Gusra Podcast." This channel features over 100 speakers discussing various topics such as sports, science, and technology, with a diverse mix of male and female voices. The statistics of this dataset are detailed in Table 12.

In addition to basic transcriptions, our manual annotation task will include multiple labels to enhance the dataset's utility. These labels will cover sentiment (e.g., positive, negative, neutral), speaker gender, and conversational context (e.g., formal, informal). By incorporating these additional labels, we aim to provide a richer dataset that can support a broader range of ASR and NLP applications such as audio sentiment analysis and dialect identification.

Table 12. CAFE-large Pseudo-labeled data

| Compressed size | 9.74GO | Gender | F: 18, M: 70 |
|---|---|---|---|
| Uncompressed size | 12.0GO | Number of utterances | 3588 |
| Total Hours | 34.58 | Audio format | Wav |
| Number of voices | 90 | Frequency | 48K |

Given our limited resources, we opted to preprocess and transcribe the data in separate stages. Initially, we employed our Preprocessing-V2 pipeline to clean the data comprehensively. Subsequently, we used the Whisper model for transcription, with a temperature setting of 0.2 to ensure accuracy. Figure 6 illustrates this phase.

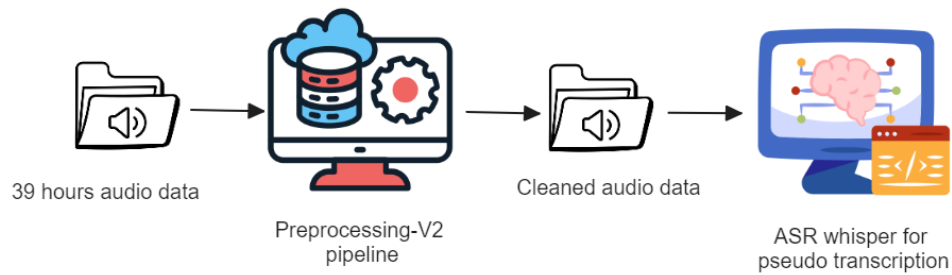

Fig. 6. Illustration of the pseudo labeling approach

This dataset forms the foundation of a larger effort to create a manually annotated corpus with detailed labels and manual segmentation. We have launched a dedicated annotation website, based on the open-source repository Audino on GitHub [16]. Involving students, and teachers, and recruiting additional personnel, we aim to enhance the dataset's quality through comprehensive manual review and annotation. Upon completion, we plan to publicly share this data, significantly contributing to the research community by providing essential resources for developing ASR systems capable of handling Algerian code-switching involving French, Arabic, and English.

The pseudo-labeled data is already publicly available in this link [7], and the ASR pseudo-labeling framework used for it is available in this link [8]. For the manual transcription project, we plan to upload the data to our website in batches, optimizing the use of material and computing resources. We are committed to sharing this data with the research community upon the completion of the annotation process, thus providing a robust foundation for future advancements in ASR systems for the Algerian dialect.

In summary, our efforts to develop a pseudo-labeled corpus for Algerian dialect, French, and English code-switching mark a significant step towards addressing the lack of specialized data in this area. By leveraging community engagement and open-source tools, we aim to create a valuable resource for researchers and developers working on advanced ASR systems.

## 8 Conclusion and Future work

In conclusion, this study successfully addresses the challenges of ASR for the Algerian dialect, particularly in spontaneous and code-switched contexts. The creation of the CAFE dataset, featuring diverse linguistic patterns and real-world speech dynamics, marks a significant advancement in this under-researched area. Our exploration of Whisper-based models, coupled with advanced prompt engineering and decoding strategies, has led to notable improvements in transcription accuracy.

---

[7] Data download link available after acceptance.
[8] The GitHub repository link will be available after acceptance

Initially, our benchmarks using the PromptingWhisper implementation with the large-v2 model showed a high Mixed Error Rate (MER) and Character Error Rate (CER) of 0.73. By upgrading to the Whisper large-v3 model, we achieved reduced MER and CER values of 0.64 and 0.66, respectively. However, it was observed that multilingual and bilingual prompting did not significantly impact the results on the CAFE-small dataset, as all prompt experiments yielded similar outcomes, even when changing the order of the language tokens in the prompt (e.g., ar-en-fr gives the same results as fr-en-ar). Further improvements were realized with advanced decoding techniques and a sophisticated preprocessing pipeline, resulting in MER and CER values of 0.310 and 0.329. These findings demonstrate substantial improvements in ASR performance.

Additionally, the development of a preprocessing layer further enhances the model's performance, making it more adept at handling the intricacies of multilingual and code-switched speech. Future work will focus on expanding the CAFE dataset to include more varied sources and larger speaker pools to enhance its representativeness. Additionally, we plan to refine the preprocessing techniques and explore more sophisticated machine learning models to further improve ASR performance. We also aim to integrate manual annotation and pseudo-labeling approaches to enrich the dataset's quality and reliability. By sharing our dataset and findings with the research community, we hope to foster further advancements in ASR systems for the Algerian Arabic Dialect and contribute to the broader field of multilingual speech recognition.

Ethics Statement

We thank the Gusra Podcast team for their permission to use their data. All data collection and annotation efforts have been conducted in accordance with ethical guidelines, ensuring respect for privacy and intellectual property rights. We are committed to maintaining transparency and integrity throughout our research process.

Acknowledgments

We would like to express our sincere gratitude to Dr. Abdelwahab Heba for his invaluable feedback and advice throughout this work.